# Mount Rainier and Liberty Cap Elevation Survey 2025, Permit MORA-2025-SCI-0005

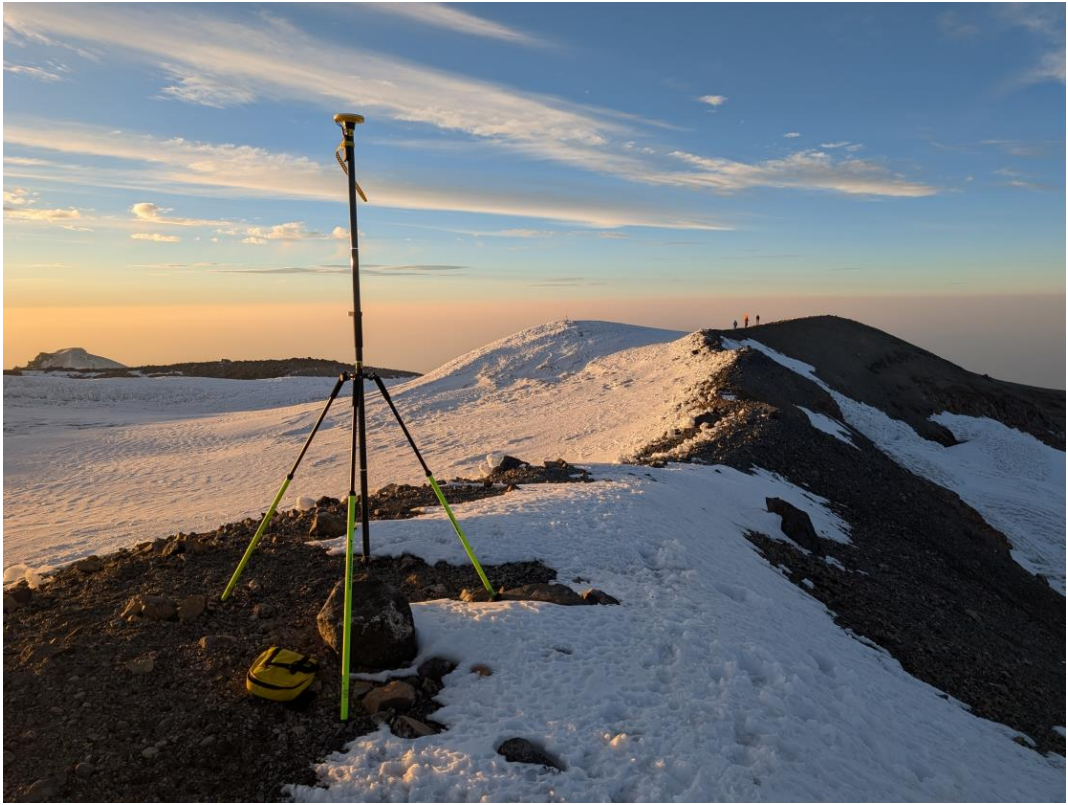

Mount Rainier summit (foreground) and Columbia Crest (background), Mount Rainier National Park, Washington, Aug. 24, 2025.

Eric Gilbertson[1], Larry Signani[2,] Branden Joy[3], Daniel McGrath[4], Darin Loucks[5], Ethan O'Connor[6], Shannon Cheng[7], Scott Hotaling[8]

[1] Seattle University, Seattle, WA
[2] Enumclaw, WA
[3] Bothell, WA
[4] Colorado State University, Fort Collins, CO
[5] Seattle, WA
[6] Seattle, WA
[7] Seattle, WA
[8] Utah State University, Logan, Utah

# Abstract

In 2024, we discovered that Columbia Crest, the historical summit of Mount Rainier, was no longer the highest point on the mountain. Instead, a point 133 m (436 ft) to the south along the southwest rim (SW Rim) was determined to be the new highest point on Mount Rainier. The Columbia Crest icecap melted down 6.64 m (21.8 ft) since 1998. A nearby peak, Liberty Cap, melted down 8.02 m (26.3 ft) since 2007. For this report, updated elevation measurements were taken in late summer 2025 of Columbia Crest and Liberty Cap. Columbia Crest melted an additional 0.37 m (1.2 ft) and Liberty Cap melted 0.67 m (2.2 ft) over the last year. To understand how Mount Rainier's summit may continue to evolve, we used ground-penetrating radar (GPR) to measure the thickness of the Columbia Crest icecap and Liberty Cap. As of 2025, ice is approximately 2 m-5 m (6.6 ft – 16.5 ft) thick near the summit of Columbia Crest and 10 m – 13 m (32.8 ft – 42.7ft) thick near the summit of Liberty Cap. Using recent measurements by our team and other surveyors, melt rates for Columbia Crest have been -0.27 m/year (-0.9 ft/year) since 1998 ($R^2 = 0.990$) and -0.49 m/year (-1.6 ft/year) for Liberty Cap since 2007 ($R^2 = 0.998$). Based on the elevation of the nearest high-elevation visible rock, Liberty Cap will lose its status as an ice-capped peak by 2041. Assuming recent rates of change continue, it will melt to bedrock by 2047. Columbia Crest is already no longer the highest point on Mount Rainier, and it will likely melt to bedrock by 2045. As of 2025, our results indicate that the summit of Mount Rainier on the SW Rim is at an elevation of 4391.04 m (14406.3 ft $\pm$ 0.1 ft) (NAVD88).

# List of Terms or Acronyms

**NGVD29:** National Geodetic Vertical Datum of 1929

**NAVD88:** North American Vertical Datum of 1988

**LiDAR:** Light Detection and Ranging

**LSAW:** Land Surveyors Association of Washington

**USACE:** US ARMY CORPS OF ENGINEERS

**USGS:** United States Geological Survey

**GPS:** Global Positioning System

**WSRN:** Washington State Reference Network

**OPUS:** Online Positioning User Service

**CSRS-PPP:** Canadian Spatial Reference System Precise Point Positioning processing

## Acknowledgements

Funding for this project was provided by the American Alpine Club, with equipment provided by Seattle University and Trimble. Juliette Jacquemont, Peter Steele, Peter Frick-Wright, Anthony Frischling, Scott Clark, and Joshua Hathaway helped carry gear for ground surveys. We appreciate the work of Mount Rainier Climbing Rangers and guides to keep the Disappointment Cleaver and Ingraham Direct routes open so we could make it to the summit. Scott Beason, Taylor Kenyon, Sallie Beavers, and Chelsea Atkins facilitated permitting for data collection.

.

# Introduction

Mount Rainier is the tallest peak in Washington State, the most topographically prominent peak in the contiguous United States of America (USA), and the most heavily glaciated peak in the contiguous USA. Until recently, it was one of the few peaks in the contiguous USA with a permanent icecap on the summit.

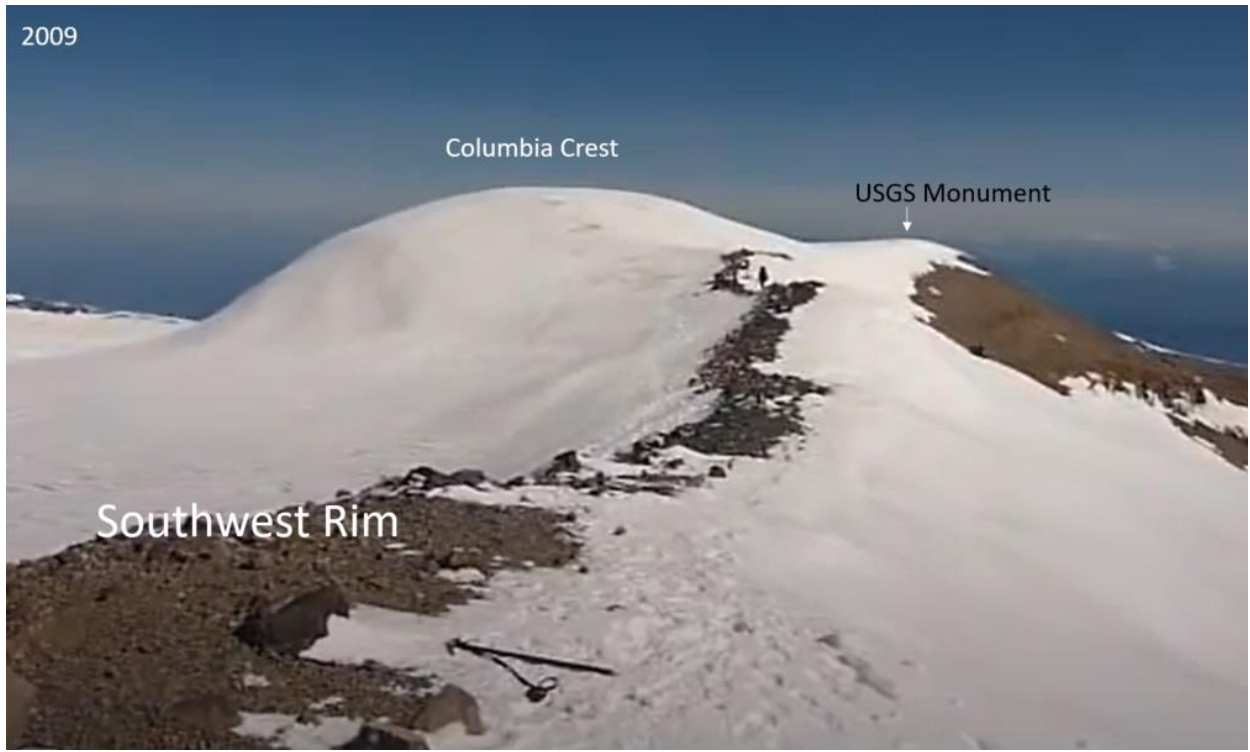

**Figure 1.** The view of Columbia Crest from the Southwest Rim in 2009 (Arhuber 2009).

The summit of Mount Rainier consists of a crater rim that partially melts down to rock every summer. However, there has historically been a permanent dome of ice on the west edge of the rim. This ice dome is referred to as Columbia Crest, and until sometime in the last 20 years, it was the highest point of the mountain (Fig. 1). Thus, the elevation of the highest point of ice on Columbia Crest in late summer has historically been considered the elevation of Mount Rainier. Columbia Crest maintained roughly constant elevation between 1956, when it was surveyed by the USGS, and 1998 when it was surveyed by the Land Surveyors Association of Washington (Signani 2000, Schrock 2011).

In the early 2000s, Columbia Crest started melting down as summit temperatures rose and yearly degree-days above freezing increased (Gilbertson et al 2025). This melting was consistent with melting of glaciers lower on Mount Rainier (Beason et al 2023), and trends observed on lower-elevation glaciers across Washington (Pelto 2025). By approximately 2014, Columbia Crest was no longer the highest point on the mountain, and Mount Rainier lost its status as one of the few ice-capped peaks in the contiguous USA. Historically, there were five in the contiguous USA, all in Washington: Mount Rainier, Liberty Cap, Colfax Peak, Eldorado Peak, and East Fury (Gilbertson et al 2025). From 1998 to 2024, Columbia Crest lost 6.64 m (21.8 ft) of elevation.

Since approximately 2014, a rock on the southwest rim of the summit crater has been Mount Rainier's true summit (Fig. 2). This location is officially unnamed but will be referred to hereafter as SW Rim.

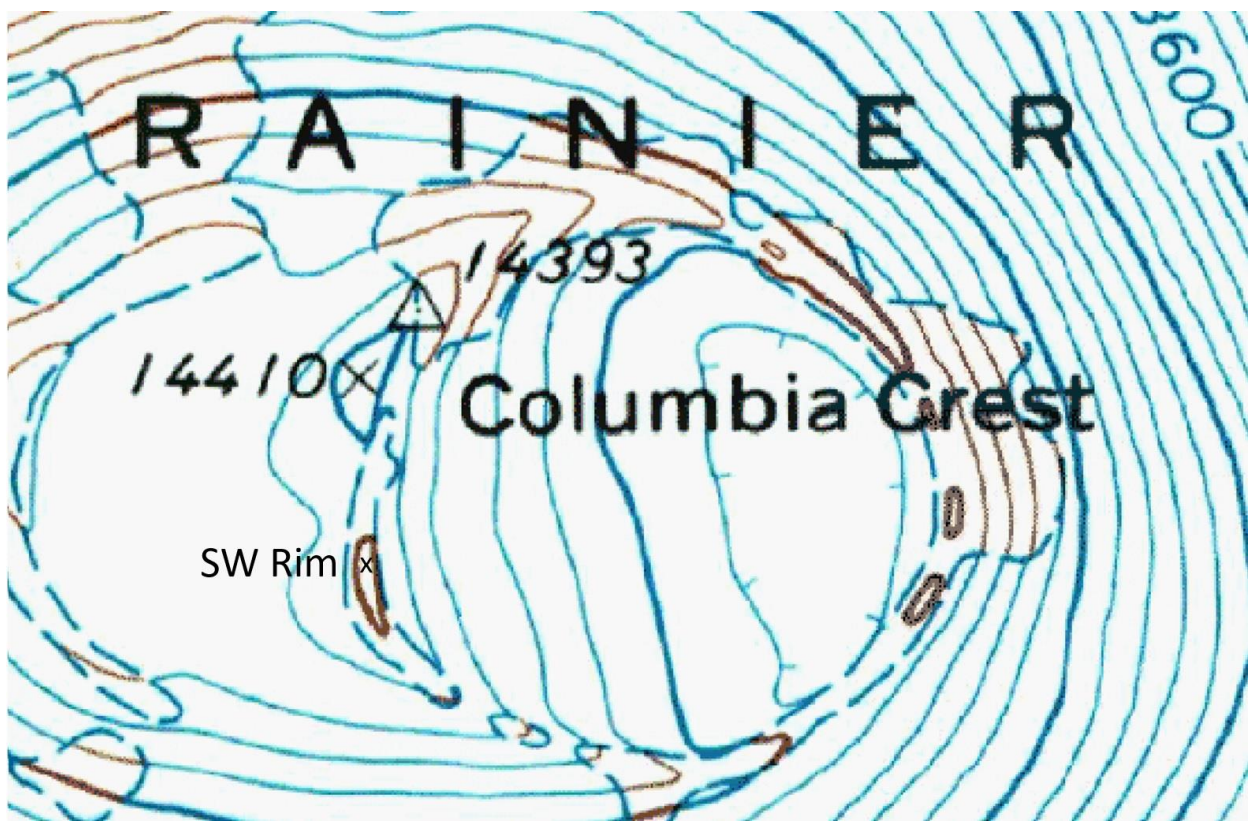

**Figure 2.** The summit region of Mount Rainier from the 1971 USGS quad (USGS 1971), with Columbia Crest and the Southwest Rim labeled. (Elevations in ft, NGVD29).

Liberty Cap, a peak approximately 1.6 km (1 mile) northwest of Columbia Crest, has historically been an ice-capped peak. It has also melted down considerably as temperatures have warmed above freezing on its summit, but its highest point is currently still ice. From 2007 to 2024, Liberty Cap lost 8.02 m (26.3 ft) of elevation (Gilbertson 2025).

Due to the history of Columbia Crest as the historic summit of Mount Rainier, there is interest in knowing how much longer it will exist before it melts to rock. To our knowledge, the thickness of the Columbia Crest icecap has not been previously measured.

Similarly, the ice thickness on Liberty Cap has never previously been measured. If the ice thickness is quantified, it provides an approach to project when Liberty Cap will melt down enough to no longer be an ice-capped peak. As of 2024, only two ice-capped peaks remained in the contiguous USA: Liberty Cap and Colfax Peak. Thus, it is significant to study how much longer Liberty Cap will remain an icecap peak.

During 2025, we sought to answer three questions: (1) how much change in elevation has occurred for Columbia Crest and Liberty Cap since 2024? (2) what is the thickness of the Columbia Crest and Liberty icecaps? And, (3) what are the seasonal snow accumulations on Columbia Crest and Liberty Cap? We also used this opportunity to further confirm our previous survey results for the new SW Rim summit of Mount Rainier.

## Methods

### Equipment

*GNSS Equipment*

Elevation measurements were taken with multiple GNSS units including Trimble DA2s, a Trimble Promark 220 with an Ashtech Antenna, a Tallysman VSP6037L, and a Trimble R980. AdirPro prism tripods with 2.00 m (6.56 ft) antenna rods were used at Paradise, Camp Muir, SW Rim, and Columbia Crest and a 0.30 m (1.0 ft) antenna rod was used at McClure Rock. A flexible-leg tripod was used at Liberty Cap to save weight. A 5x 10-arcminute Sokkia Abney level was used to measure angular inclinations and declinations between the SW Rim and Columbia Crest to further validate elevation measurements.

*GPR Equipment*

Ice thickness measurements were taken with a commercial Noggin Ultra 100 ground-penetrating radar (GPR) unit with positioning provided by a TopCon SGR-1 GNSS unit.

### Measurements

*Spring*

The maximum snow depth time of year at the Paradise SNOTEL site on the south side of Mount Rainier is generally the first week of May. This was used as the approximate date of max snow depth on Columbia Crest and Liberty Cap.

On May 6, shortly after midnight, the team left Paradise and skied and climbed to Columbia Crest by 10am. A DA2 unit was mounted on a 2.00 m (6.56 ft) tripod on Columbia Crest (Fig. 3) and data collection started. The team then continued to Liberty Cap.

A Promark unit with Ashtech antenna was mounted on a 0.30 m (1.0 ft) antenna rod at 11:30am and data logged for one hour. The team then returned to Columbia Crest by 1pm and saved 3 hours of data for Columbia Crest. The team then skied down to McClure Rock by 4:30pm and recorded one hour of data with the DA2 mounted on a 0.30m (1.0ft) tripod on the monument.

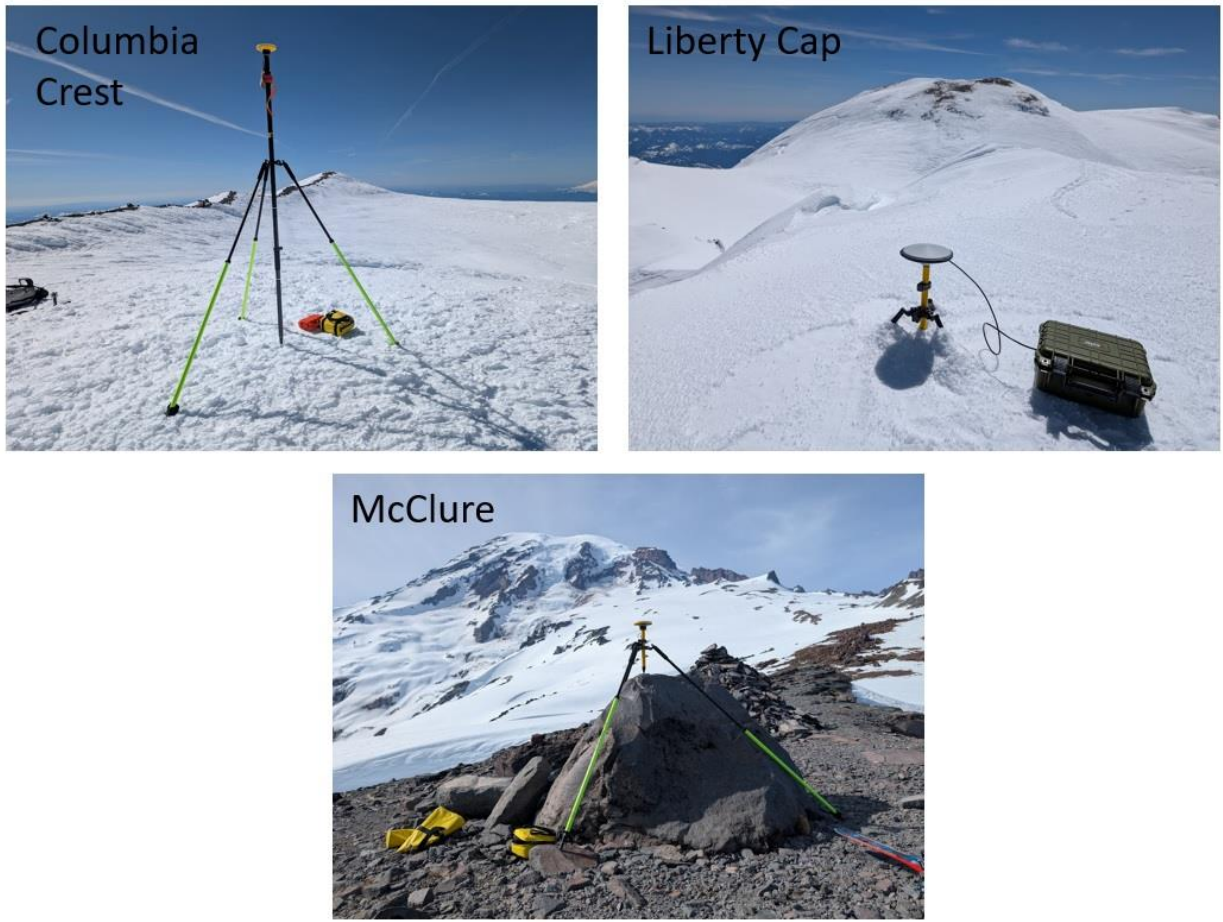

**Figure 3.** The GNSS units mounted on Columbia Crest, Liberty Cap, and McClure Rock on May 6, 2025.

*Summer*

All USGS monuments (Paradise, McClure, and Muir) were occupied on Aug 24 (Fig. 4). Measurements were taken simultaneously at all monuments and at SW Rim, Columbia Crest, and Liberty cap.

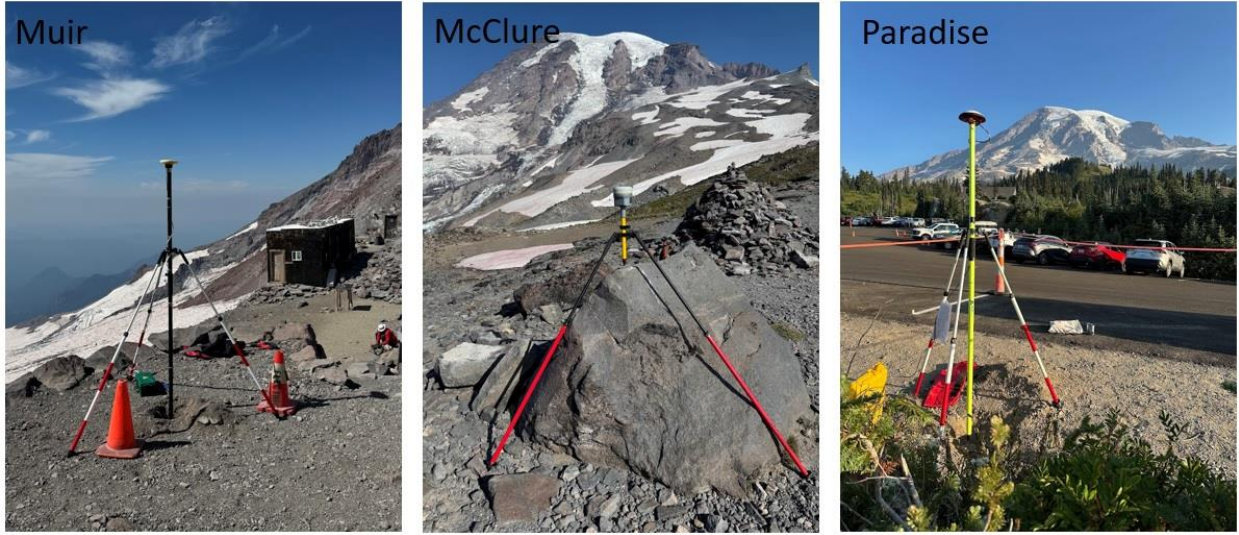

**Figure 4.** Muir, McClure, and Paradise monuments occupied Aug. 24, 2025.

The summit team left Paradise shortly after midnight Aug. 24 and reached the summit crater by noon. That morning one team member hiked to Muir, one to McClure Rock, and one remained at Paradise. At noon the summit team communicated with the lower team members over radio to start logging data at the lower monuments. Data was recorded at each lower monument for at least 3 hours (Fig. 4).

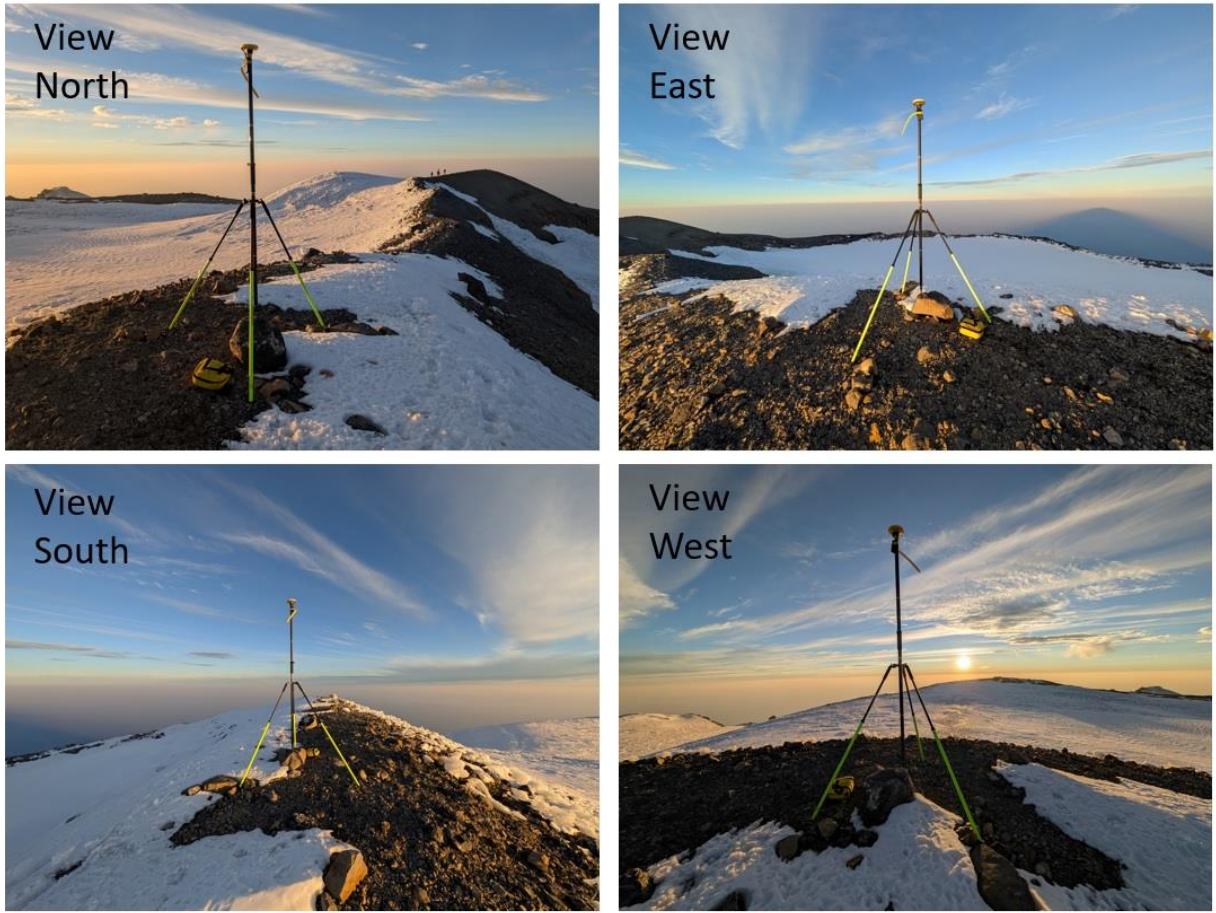

**Figure 5.** The GNSS unit mounted on the SW Rim. Views look North, East, South, and West.

At 12:30pm a Trimble DA2 unit was mounted on a 2.00 m (6.56 ft) tripod on the SW Rim and data logging commenced (Fig. 5). A second Trimble DA2 unit was mounted on another 2.00 m (6.56 ft) antenna rod on Columbia Crest and data logging began at 12:45pm. Interestingly, this year a crevasse had opened almost directly over the summit, and the antenna rod was mounted near the edge of the crevasse. This crevasse has not existed in previous years based on the experience of EG.

Abney level measurements were taken looking up from Columbia Crest to the SW Rim and also looking down from the SW Rim to Columbia Crest. These would be used to help corroborate the GNSS measurements.

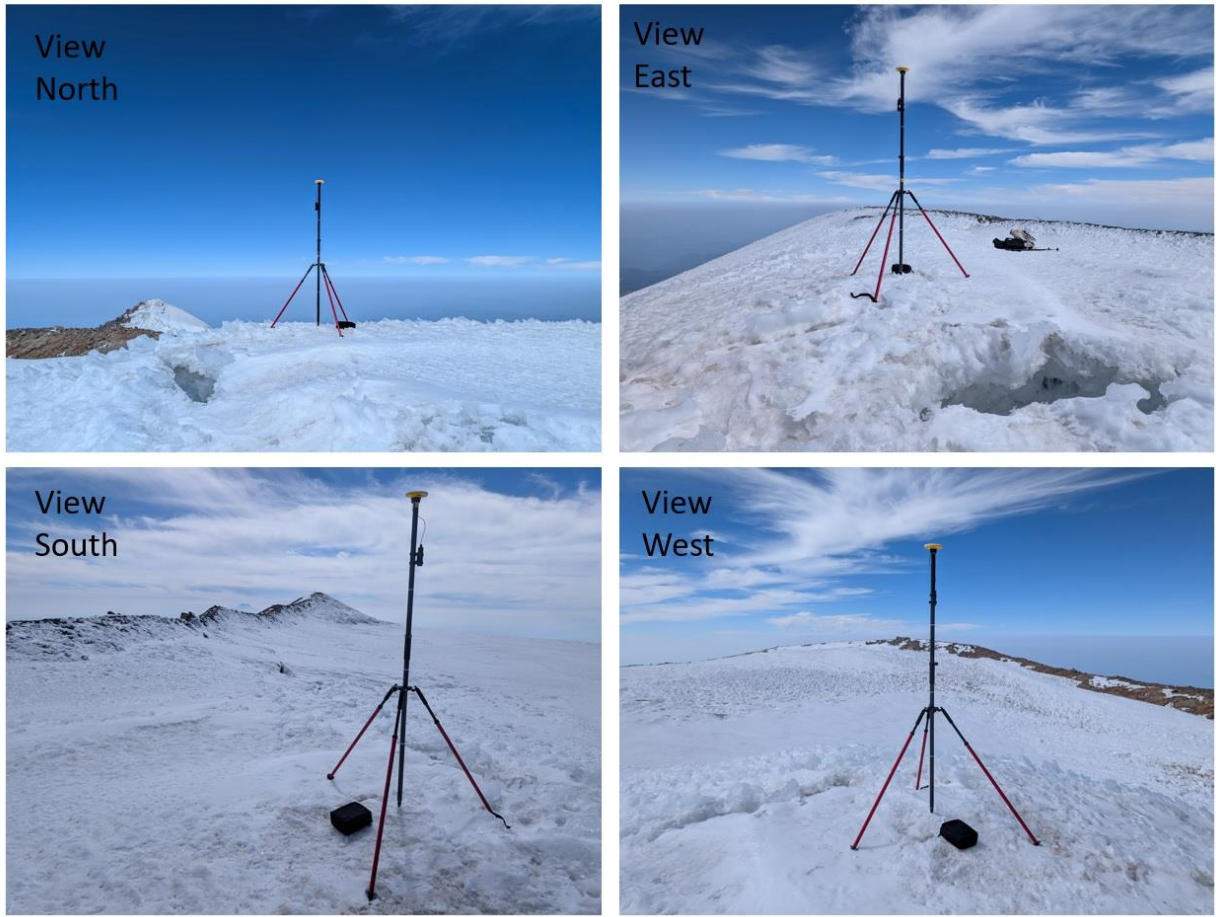

**Figure 6.** The GNSS mounted on Columbia Crest. Views look North, East, South, and West.

At 2pm, we began our GPR surveys of Columbia Crest. We recorded five linear surveys of 50-150 m per survey. One team member stayed with the SW Rim GNSS unit and another stayed with the Columbia Crest GNSS unit.

Three members continued to Liberty Cap with the GPR and one remaining GNSS unit. At 5:15pm the Liberty Cap team reached the summit. The summit was covered in 1 m (3.3 ft)-tall penitentes, which did not exist on the summit in late summer 2024. The GNSS unit was mounted at the same coordinates as were measured in 2024, on the highest point that was not a penitente (Fig. 7).

We performed a 20 m GPR survey of Liberty Cap beginning on the summit then around to the east and back in a clockwise circle. A longer GPR survey of Liberty Cap was not feasible due to the penitentes. Elevation data was logged for 45 minutes, then the team packed up the equipment at 6:30pm and returned to Columbia Crest. Seven hours of data were logged for Columbia Crest and three hours for SW Rim.

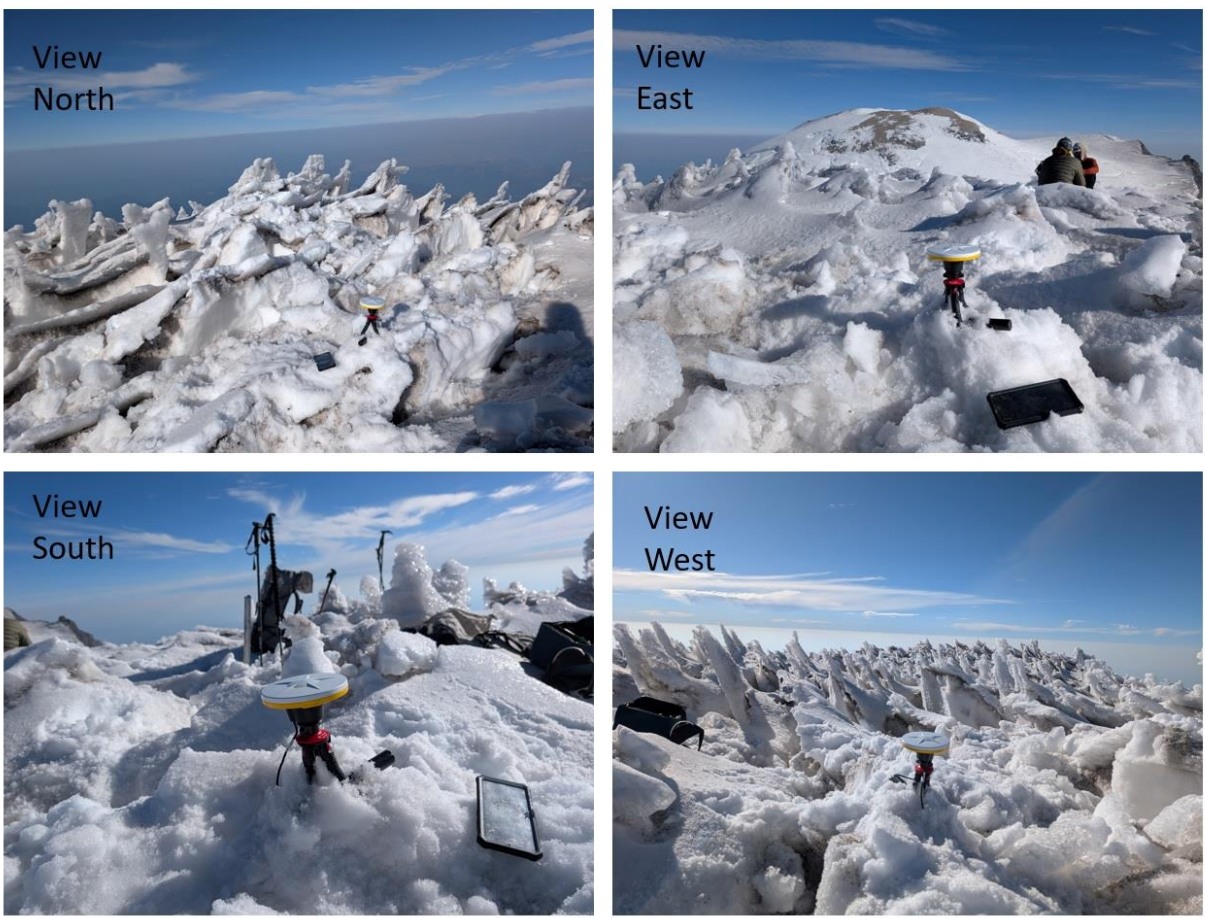

**Figure 7.** The GNSS mounted on Liberty Cap. Views look North, East, South, and West.

**Data Processing**

*GNSS Processing*

Data for summer measurements were processed with Trimble Business Center (TBC 2025), Online Positioning User Service (OPUS 2025), OPUS Projects (OPUS Projects 2025), Canadian Spatial Reference System Precise Point Positioning processing (CSRS-PPP 2025), Trimble RTX (Trimble 2025), and AUSPOS Online GPS Processing Service (Jia 2014). We report the OPUS results but note that the TBC, OPUS Projects, CSRS-PPP, Trimble RTX, and AUSPOS methods gave similar, consistent values.

Data for spring measurements were processed with OPUS.

Results will be reported in meters and US Survey feet in NAD83 (2011) Epoch 2010, NAVD88 (Geoid 18). This is the current standard datum output by OPUS.

For elevation measurements of Columbia Crest, we fit a linear model to the portion of elevation measurements between 1998 – 2025, when Columbia Crest has been melting down. For Liberty Cap we fit a linear model to the portion of elevation measurements between 2007–2025 when Liberty Cap has melted down.

*GPR Processing*

GPR surveys were processed in ReflexW with the following processing steps: i) dewow filter to remove low-frequency noise, ii) time-zero correction to align the signal transmission with the snow surface, and iii) trace resampling to ensure equidistant sampling intervals along the profile. We applied a radar velocity of 0.17 m/ns to convert measured two-way travel times to manually interpreted bedrock surfaces to estimate ice thicknesses.

# Results

**Spring Results**

A summary of all results from spring measurements can be found in Table 1.

Table 1: Spring measurements.

| Location | Elevation m (ft) | Sigma m (ft) | Lat | Lon |
|---|---|---|---|---|
| Columbia Crest | 4388.00 (14396.3) | 0.04 (0.13) | 46° 51' 10.55250" | -121° 45' 38.20475" |
| Liberty Cap | 4298.75 (14103.5) | 0.02 (0.05) | 46° 51' 46.57925" | -121° 46' 28.58167" |
| McClure Rock | 2252.62 (7390.5) | 0.02 (0.05) | 46° 48' 31.07214" | -121° 43' 22.36735" |

Liberty Cap was found to have accumulated 6.6ft of snow between Sept. 21, 2024 and May 6, 2025 (Fig. 8). Columbia Crest accumulated only 0.5ft of snow (Fig. 9). The McClure Rock monument measurement was consistent with the 2010 measurement of 7389.78ft, which corroborates the summit measurements.

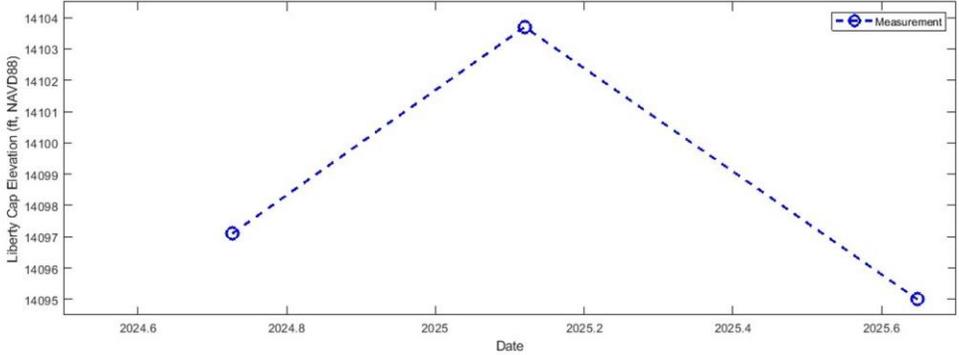

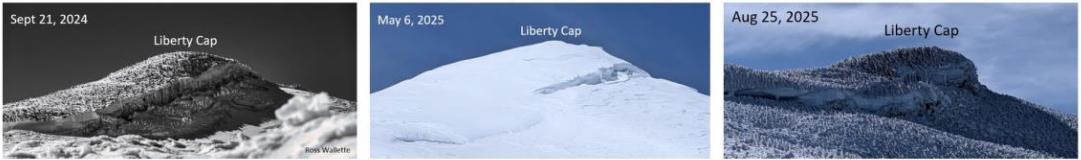

**Figure 8.** Seasonal elevation variation of Liberty Cap from Sept. 2024 to May 2025 to Aug. 2025.

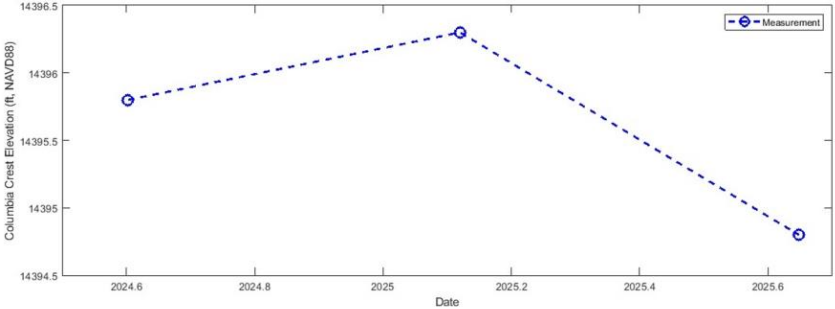

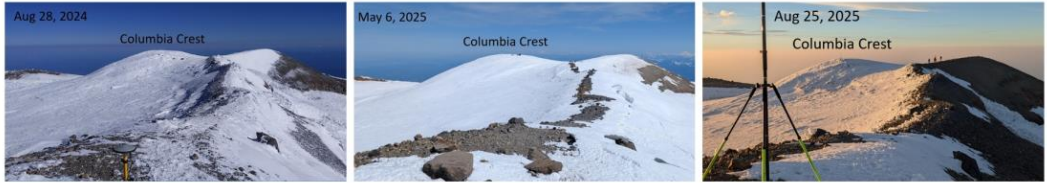

**Figure 9.** Seasonal elevation variation of Columbia Crest from Sept 2024 to May 2025 to Aug. 2025.

**Summer Results**

*GNSS Results*

A summary of summer measurements can be found in Table 2.

**Table 2:** Summer GNSS positioning measurements of monuments and upper mountain locations.

| Location | Elevation m (ft) | Sigma m (ft) | Lat | Lon |
| --- | --- | --- | --- | --- |
| SW Rim | 4391.04 (14406.3) | 0.02 (0.07) | 46° 51' 06.21562" | -121° 45' 37.43064" |
| Columbia Crest | 4387.47 (14394.6) | 0.02 (0.07) | 45° 51' 10.553091" | -121° 45' 38.20381" |
| Liberty Cap | 4296.13 (14094.9) | 0.02 (0.07) | 46° 51' 46.50016" | -121° 46' 28.75120" |
| Muir | 3076.29 (10092.8) | 0.09 (0.28) | 46° 50' 07.81199" | -121° 43' 56.44798" |
| McClure | 2252.15 (7389.8) | 0.02 (0.07) | 46° 48' 31.07171" | -121° 43' 22.36721" |
| Paradise | 1644.09 (5394.0) | 0.01 (0.02) | 46° 47' 06.91811" | -121° 44' 03.82096" |

The summit of Mt Rainier on the SW Rim was measured to be 4391.04 m $\pm$0.04m (14406.3 ft $\pm$0.1 ft) (95% confidence interval). This is consistent with the reported 2024 measurements of 4391.01 m $\pm$0.03 m (14406.2 ft $\pm$0.1 ft) (95% confidence interval errors).

Columbia Crest was measured to be 4387.47 m $\pm$0.04 m (14394.6 ft $\pm$0.1 ft) (95% confidence interval errors). This means Columbia Crest lost 0.37m (1.2ft) of elevation between Aug. 28, 2024 and Aug 24, 2025.

Thus, as of Aug. 2025, the SW Rim is 3.57 m (11.7 ft) taller than Columbia Crest and is the highest point on Mount Rainier.

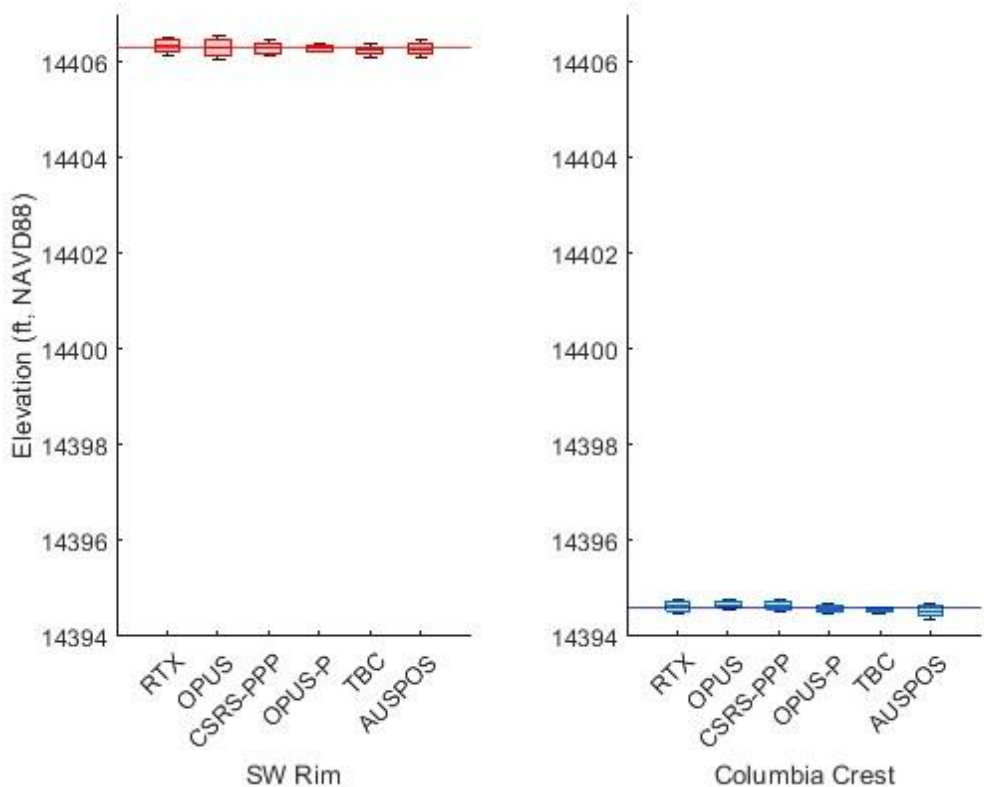

**Figure 10.** Box-and-whisker plots comparing elevations of SW Rim and Columbia Crest for different processing methods. Boxes are centered at mean measurements with box edges at +/- one sigma and whisker edges at $\pm$ two sigma error bounds.

All GNSS measurement processing methods were consistent that the SW Rim is higher than Columbia Crest. This can be seen in box-and-whisker plots (Fig. 10). Measured elevations are over 50-sigma apart, meaning it is essentially statistically certain that the SW Rim is higher than Columbia Crest.

Abney level measurements found an angular inclination of 1.5 degrees $\pm$5 arcminutes up from Columbia Crest to the SW Rim and 1.5 degrees $\pm$5 arcminutes down from the SW Rim to Columbia Crest. The horizontal distance is 133 m (436 ft) between them, which means the measured height difference from basic trigonometry is 3.47 m $\pm$0.02 m (11.4 ft $\pm$0.8 ft). This is consistent with the 11.7 ft height difference measured by the GNSS units.

Liberty Cap was measured to be 4296.13 m $\pm$0.04 m (14094.9 ft $\pm$0.1ft). Liberty Cap melted down 0.67 m (2.2ft) from Sept. 21, 2024 to Aug. 24, 2025. It still had ice at its highest point, so it is still an ice-capped peak.

*GPR Results*

Five data lines were measured on Columbia Crest by the Noggin GPR device, and one representative line is shown (Fig. 11). This line was 130 m long and started northeast of the summit, then followed a clockwise spiral ending on the summit.

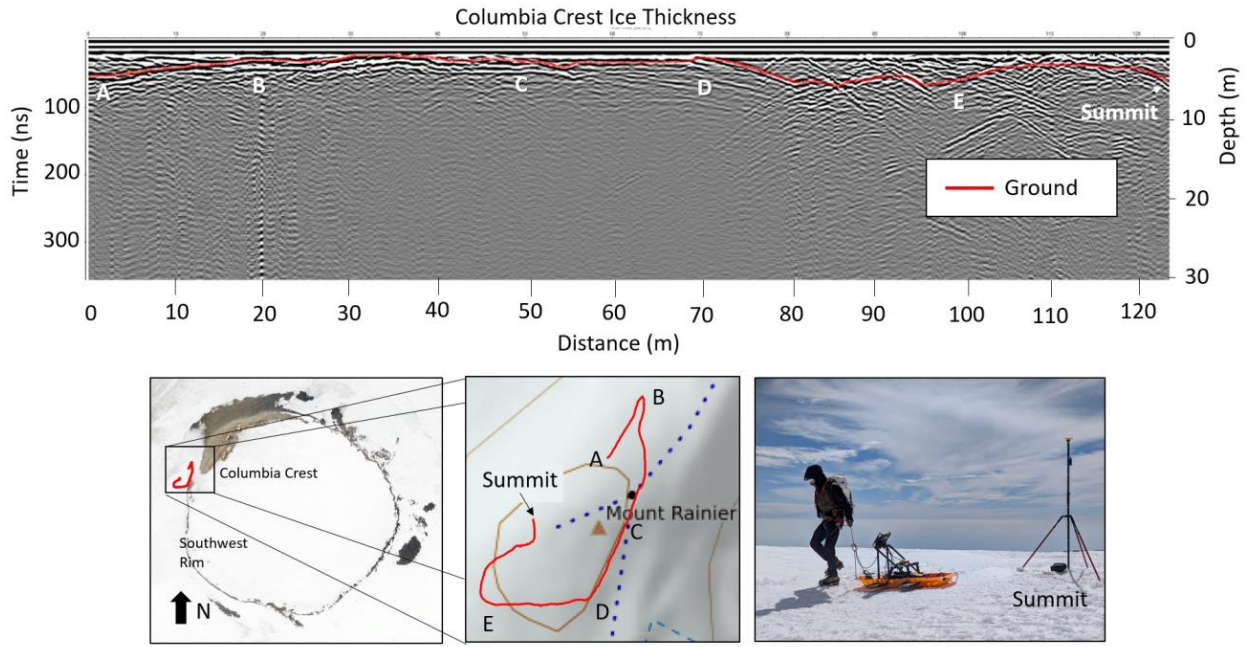

**Figure 11.** Ground-penetrating Radar depth measurements (top) for 130 m horizontal-distance line on Columbia Crest (bottom left, bottom center). The line ended on the summit (bottom right).

The surveyed area of Columbia Crest has an ice thickness between 1 m–5 m (3.3 ft – 16.5 ft). The ice is 5 m (16.5 ft) thick on the exact summit. Subtracting the ice thickness from the measured elevation of the Columbia Crest summit means the underlying bedrock has a measured elevation of 4382.44 m (14378.1 ft).

We fit a linear model to the portion of elevation measurements between 1998 – 2025, when Columbia Crest has been melting down (Fig. 12). This fit was highly linear, with an $R^2$ of 0.990 and slope -0.9 ft/year. Columbia Crest has been losing -0.9 ft per year of elevation since 1998. At this rate, assuming the same melt rate persists in future years, the Columbia Crest icecap will fully melt by approximately 2045.

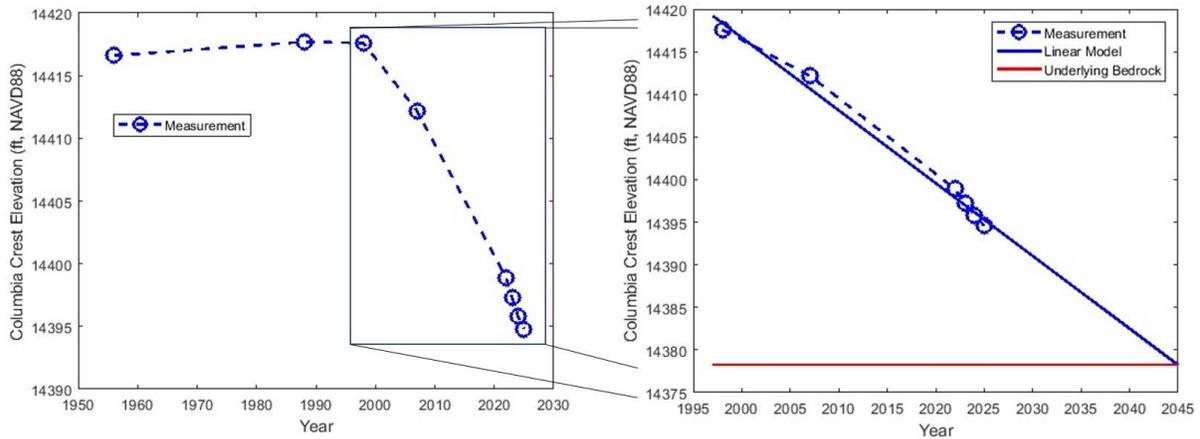

**Figure 12.** Columbia Crest elevation over time between 1956 – 2025 (left), and linear model fit for time between 1998-2025 (right) with underlying bedrock elevation shown.

For Liberty Cap two GPR data lines were collected, and the longest is shown. This line was 11 m (36.1 ft) long, started on the summit, and moved in a clockwise direction (Fig 13). The surveyed area of Liberty Cap has an ice thickness between 10.6 m – 13.0 m (34.8 ft – 42.7 ft). The ice is 10.6 m (34.8 ft) thick on the exact summit. Subtracting the ice thickness from the measured elevation of Liberty Cap means the underlying bedrock has a measured elevation of 4285.52 m (14060.1 ft)

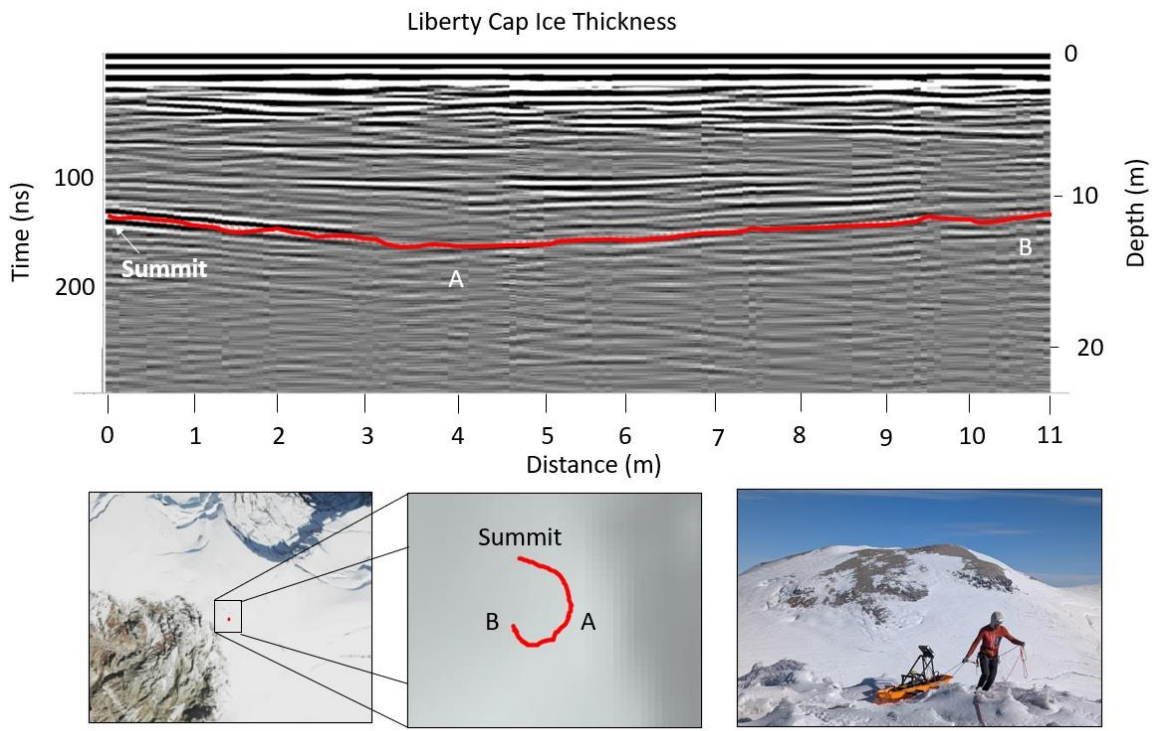

**Figure 13.** Ground-penetrating Radar depth measurements (top) for 11m horizontal-distance line on Liberty Cap (bottom left, bottom center). The line started on the summit (bottom right).

The elevation of the highest visible rock outcrop on Liberty Cap has been found based on LiDAR measurements (Gilbertson 2025) to be at an elevation of 14069.4 ft. If Liberty Cap melts below this elevation it will no longer be an icecap peak.

Elevation measurements for Liberty Cap are plotted for data between 1956 and 2025 (Fig 14). A linear model is fit to the data between 2007 and 2025 when Liberty Cap has been melting down. This model is also highly linear, with an R squared value 0.998 and a slope of -0.49 m (-1.6 ft) per year.

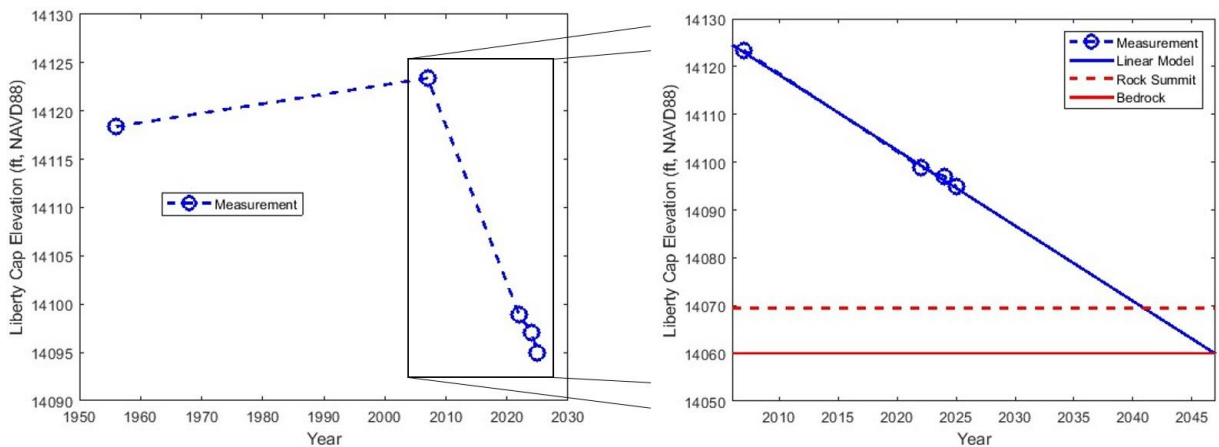

**Figure 14.** Liberty Cap elevation over time between 1956 – 2025 (left), and linear model fit for time between 1998–2025 (right) with highest visible rock elevation shown.

This means Liberty Cap has been melting down at an average rate of -0.49 m (-1.6 ft) per year since 2007. If this melt rate continues, the summit will melt below the elevation of the highest visible rock by 2041, and Liberty Cap will no longer be an icecap peak. By 2047 Liberty Cap will melt down to bedrock.

## Discussion

The true summit of Mt Rainier is the SW Rim, at 4391.04 m $\pm 0.04$ m (14,406.3 ft $\pm 0.1$ ft) (95% confidence interval). This has been the summit since approximately 2014. It was still measured as the highest point in 2025, consistent with 2024 results.

Columbia Crest began melting down in 1998 and has been melting at a surprisingly linear rate (R squared 0.990), losing 0.027 m (0.9 ft) each year very consistently. It is unclear why this rate is so linear and not accelerating. Future work could investigate in detail if the modeled temperatures on the summit have similarly increased linearly. Though, without direct temperature measurements on the summit it is hard to know for certain what past temperatures were. If temperature could be measured directly on the summit for one summer season, it may be possible to calibrate temperature models from past years to get a more accurate idea of past summit temperatures.

Columbia Crest was measured to be only 5 m (16.5 ft) thick at the summit in 2025. At current melt rates, it will melt to bedrock by 2045 (Fig 15).

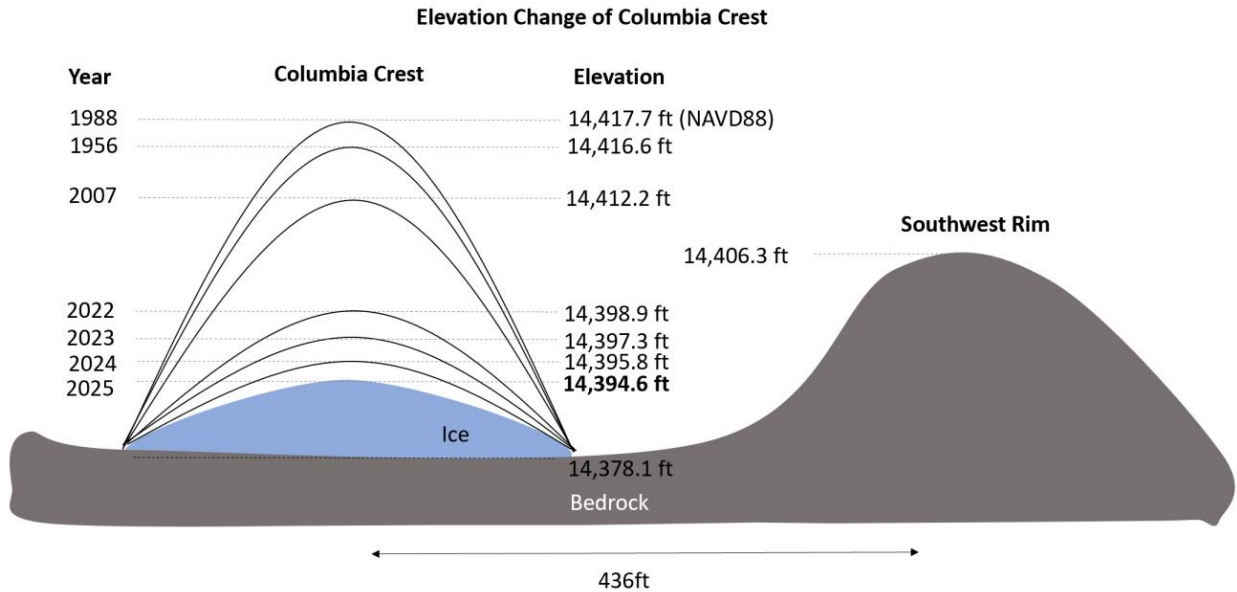

**Figure 15.** Schematic diagram of Mt Rainier SW Rim and Columbia Crest elevations over time.

Columbia Crest only accumulated 0.15 m (0.5 ft) of snow between late summer 2024 and spring 2025, at the approximate maximum snow depth time of year. Certainly much more snow than this falls on the summit. One possible explanation is that it all gets blown off and does not accumulate. The accumulation in May appeared to be more rime than snow, which may be because rime can form in high wind while snow might get blown off.

This leads to a potential explanation of the seasonal and long-term behavior of the elevation of Columbia Crest. Before 1998, Columbia Crest appeared to be at a steady-state elevation based on all measurements. In the winter and spring snow did not accumulate because it was mostly blown off. In the summer Columbia Crest did not melt down appreciably because it rarely got above freezing (see Gilbertson 2025).

However, beginning in the late 1990s or early 2000s, the temperature at the summit began getting above freezing a significant amount. In each of these years it still did not accumulate much snow in the winter/spring, but it began melting down approximately 0.27 m (0.9 ft) every summer. That trend has continued through 2025.

It is unknown exactly how Columbia Crest formed, and why or when snow might have accumulated in that area. If ice cores could be taken into Columbia Crest, different layers of historical accumulation could be examined and dated. That could potentially reveal information about when and how fast Columbia Crest formed.

Liberty Cap started melting down around 2007 (or perhaps earlier, though no measurements exist between 1956 – 2007). It has been melting at a highly linear rate (R squared value 0.998), just like Columbia Crest has. If the current melt rates continue, it will no longer be an icecap peak by 2041 (Fig 16). As of 2025 there are only two icecap peaks remaining in the contiguous US – Colfax peak and Liberty Cap. Colfax peak will likely lose its status as an icecap peak within a few years at current melt rates (a paper about this is in preparation). This means Liberty Cap will likely be the last remaining icecap peak in the contiguous USA soon, and by 2041 there will no longer be any icecap peak in the contiguous USA. (Note: Columbia Crest now counts as a sub-peak of Mt Rainier because it has such low prominence).

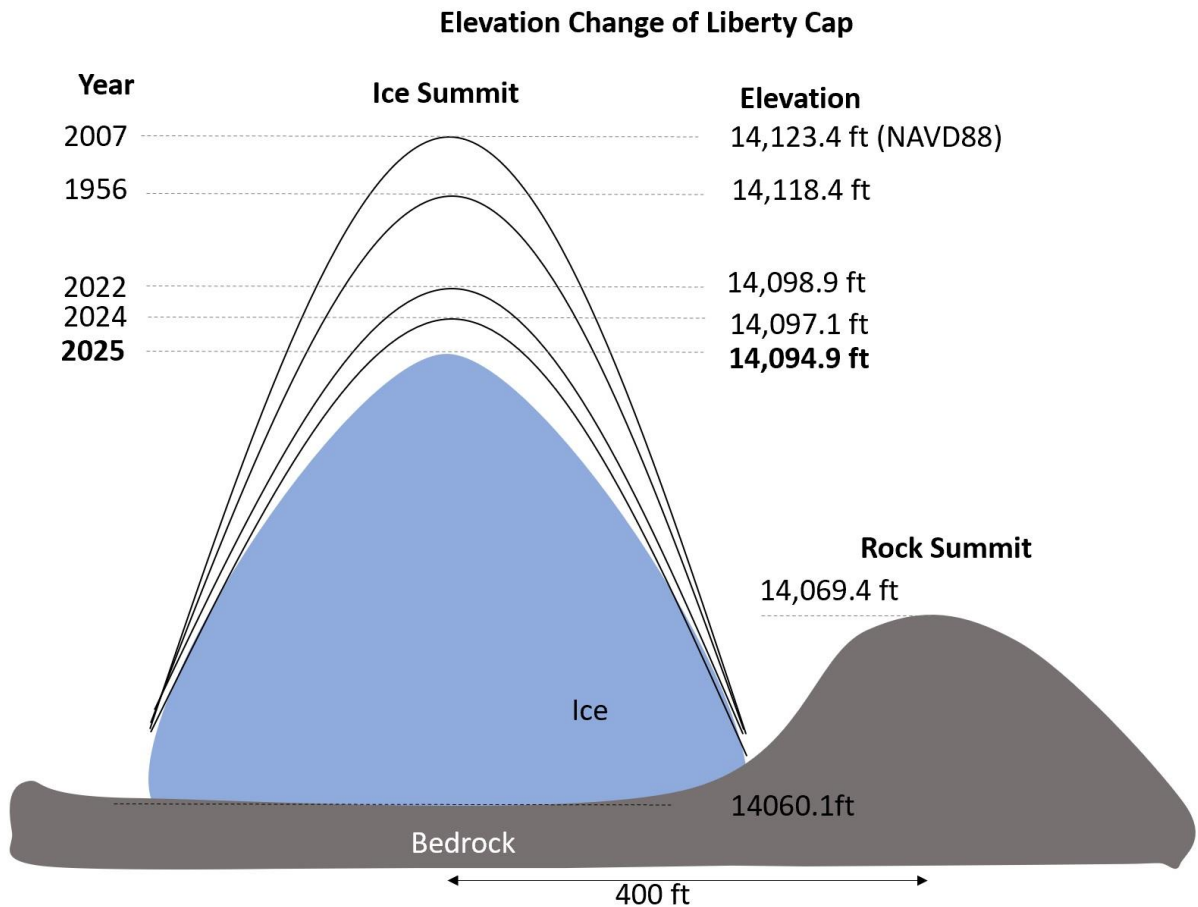

**Figure 16.** Schematic diagram of Liberty Cap elevations over time.

Interestingly, the elevation loss rate of Liberty Cap is almost twice the rate of Columbia Crest. This phenomenon of different elevation loss rates has been documented on other pairs of nearby icecap peaks around the world, such as Pico Bolivar and Pico Colon in Colombia (Gilbertson et al, in review). It is possible that differences in icecap peak shape have a strong affect on their elevation loss rates because this affects wind scouring and winter accumulation.

Also, Columbia Crest is about 300ft taller than Liberty Cap, which would result in about 1 degree F colder temperature (assuming a standard lapse rate of 3.5°F per 1000 ft). It is possible freezing levels are often at Liberty Cap elevation but just below Columbia Crest elevation. More detailed temperature data would be needed to understand this.

Liberty Cap accumulated 6.6 ft of snow between Sept. 21, 2024 and May 6, 2025. This was over 10 times the accumulation on Columbia Crest, even though the peaks are only 1.6 km (1 mile) apart horizontally. This is possibly because wind affects are less pronounced on Liberty Cap, and less snow gets blown off the top. Columbia Crest is at a higher elevation and is thus likely subject to stronger winds.

## Conclusion

Columbia Crest melted down 0.37 m (1.2 ft) between late summer 2024 and late summer 2025 and had an ice thickness of 5 m (16.5 ft) at the summit in 2025. Melt rates were highly linear, and if the current rates continue into the future, it will melt to bedrock by 2045. Only a minimal amount of snow (0.15 m/0.5 ft) accumulated on Columbia Crest in winter/spring 2025.

Liberty Cap melted down 0.67m (2.2ft) between late summer 2024 and late summer 2025. If current highly linear melt rates continue it will no longer be an icecap peak by 2041, and will melt to bedrock by 2047.

The Southwest Rim is still the highest point on Mt Rainier, with an elevation of 4391.04m $\pm$0.01 m (14,406.3 ft $\pm$0.1 ft) (NAVD88).

## Data Availability

Raw measurement files are available at:
https://drive.google.com/drive/u/0/folders/1sZM3rWVBY5cJwaRu41GiaI8cGl72rGw9

## Literature Cited

Arhuber, Rainier summit view June 27, 2009, https://www.youtube.com/watch?v=qDQ6jqAs2h0

Beason SR and Others. 2023. Changes in glacier extents and estimated changes in glacial volume at Mount Rainier National Park, Washington, USA from 1896 to 2021. Natural Resource Report. NPS/MORA/NRR—2023/2524. National Park Service. Fort Collins, Colorado. https://doi.org/10.36967/2299328

CSRS-PPP. 2024. Canadian Spatial Reference System Precise Point Positioning processing.

Gilbertson, E., Abatzoglou, J., Stanchak, K., Hotaling, S., "Rapid shrinking and loss of ice-capped summits in the western USA," Arctic, Antarctic, and Alpine Research Journal, 2025 Access at: https://www.tandfonline.com/doi/full/10.1080/15230430.2025.2572898

Gilbertson, E., Stanchak, K., Hotaling, S., "Contemporary Shrinking of Colombia's Highest Mountains: Pico Simón Bolivar and Pico Cristóbal Colón," Geografiska Annaler: Series A, Physical Geography, (in review)

Jia, Minghai, Dawson, John, Moore, Michael, "AUSPOS: Geoscience Australia's On-line GPS Positioning Service," *Proceedings of the 27th International Technical Meeting of the Satellite Division of The Institute of Navigation (ION GNSS+ 2014)*, Tampa, Florida, September 2014, pp. 315-320

Mount Rainier National Park, 2024 https://www.nps.gov/mora/index.htm

NGS Coordinate Conversion and Transformation Tool (NCAT), 2025 https://www.ngs.noaa.gov/NCAT/

Online Positioning User Service (OPUS), NOAA, 2024 https://geodesy.noaa.gov/OPUS/

OPUS Projects, Version 5.3, 2025, https://geodesy.noaa.gov/OPUS-Projects/OpusProjects.shtml

Pelto, M.S.; Pelto, J. The Loss of Ice Worm Glacier, North Cascade Range, Washington USA. *Water* **2025**, *17*, 432. https://doi.org/10.3390/w17030432

Signani, L., "The Height of Accuracy," July 19, 2000, Point of Beginning, https://archive.ph/IVhw#selection-1113.5-1113.26

Schrock, G. "Rainier – The Unforgiving Mountain," Jan 27, 2011, The American Surveyor, https://amerisurv.com/2011/01/27/rainier-the-unforgiving-mountain/

Trimble RTX Post Processing, Surveytools, (2025) https://surveytools.trimbleaccess.com/gnssprocessor

Trimble Business Center, Version 5.50, accessed Nov 25, 2025

U.S. Geological Survey, 20140926, USGS Lidar Point Cloud WA_MOUNTRANIER_2007_2008 WA_MountRanier_2007-2008_000419: U.S. Geological Survey.

U.S. Geological Survey, 20220901, USGS Lidar Point Cloud WA_MOUNTRANIER_2022 WA_MountRanier_2022: U.S. Geological Survey.

USGS, "What is Lidar and Where Can I Download It?", 2024 https://www.usgs.gov/faqs/what-lidar-data-and-where-can-i-download-it

USGS National Map Downloader, 2024. https://apps.nationalmap.gov/downloader/ U.S. Geological Survey: Mt. Rainier West quadrangle, Washington [map]. Photogrammetry 1970. Field checked 1971. 1:24,000. United States Department of the Interior, USGS, 1971

WSRN. 2024. Washington State Reference Network, http://www.wsrn.org/

USGS National Map Downloader, 2024. https://apps.nationalmap.gov/downloader/ U.S. Geological Survey: Mt. Rainier West quadrangle, Washington [map]. Photogrammetry 1970. Field checked 1971. 1:24,000. United States Department of the Interior, USGS, 1971

WSRN. 2024. Washington State Reference Network, http://www.wsrn.org/